\documentclass[english,10pt,a4paper]{article}
\usepackage[utf8x]{inputenc}
\usepackage{babel}
\usepackage{graphics}
\usepackage{graphicx}
\usepackage{epsfig}
\usepackage{float}
\usepackage{amsmath}
\usepackage{amssymb}

\begin{document}

\title{Parity Realization in Lattice QCD with Ginsparg-Wilson
  Fermions}

\author{V. Azcoiti, E. Follana, A. Vaquero,\\
	Universidad de Zaragoza\\
	G. Di Carlo\\
	Laboratorio Nazionale di Gran Sasso\\
	}

\date{\today}
\maketitle

\begin{abstract}
The Vafa-Witten arguments on the realization of parity and flavour
symmetries in the QCD vacuum do not apply to two lattice
regularizations of QCD which are able to reproduce the chiral anomaly:
Wilson fermions and Ginsparg-Wilson fermions. We show here how using
the last regularization one can get, from first principles, that the
more standard order parameters for these symmetries take a vanishing
vacuum expectation value for any number of flavours, if quarks have a
non-vanishing mass.
\end{abstract}

\section{Introduction}
Since the very beginning of the formulation of QCD as the gauge theory
which describes the strong interaction between elementary particles,
the understanding of the realization of symmetries in the QCD vacuum
has been an important issue. During the 80’s, Vafa and Witten gave
arguments against spontaneous breaking of parity \cite{Vafa1} and
vector-like global symmetries \cite{Vafa2} in vector-like theories
such as QCD, and in principle, this seemed to settle the matter;
however, these arguments were not as useful as expected. Indeed some
years after the publication of \cite{Vafa1}, several articles appeared
\cite{ag,Ji,Crom,av}, calling into question the validity of the
paper. The fact that the issue is still open twenty years after the
publication of the first paper is indicative of the complexity of the
subject. Regarding vector-like symmetries such as flavour or baryon
number conservation, it must be remarked that the Vafa-Witten theorem
\cite{Vafa2} is not applicable neither to the Ginsparg-Wilson
regularization nor to one of the most used fermionic regularizations
on the lattice for QCD, i.e., Wilson fermions. In the Ginsparg-Wilson
case the theorem does not apply because even if the integration
measure is positive definite, the other essential ingredient in the
proof in \cite{Vafa2}, the anticommutation of the Dirac operator with
$\gamma_5$, is not realized. For the case of Wilson fermions neither
of the two assumptions in \cite{Vafa2}, positivity of the integration
measure and anticommutation of the Dirac operator with $\gamma_5$, are
fulfilled, the first of the two failing for an odd number of
flavours. Indeed there exists a region of the parameters space where
parity and flavour symmetries are spontaneously broken: the well known
Aoki phase \cite{Aoki,Aoki2}, and even a more complex phase structure
for lattice QCD with Wilson fermions has been recently suggested
\cite{yo2,Sharpe}. In the end, a theoretical proof of the realization
of symmetries of QCD is still lacking.

The standard wisdom is that Vafa and Witten theorems fail, when
applied to Wilson fermions, due to the existence of exceptional
configurations which have a non-vanishing weight in the Aoki
phase. Outside this phase, and in particular in the physical region
near the continuum limit, the exceptional configurations would be
suppressed and then parity and flavour symmetries would be restored in
the QCD vacuum.  Following this wisdom, it would be very convenient to
choose a ‘small eigenvalue free’ regularization of QCD. It happens
that Ginsparg-Wilson fermions fulfill this requirement.

In this paper we will show how the probability distribution function
(p.d.f.)  of all the standard fermion bilinear order parameters
\cite{pdf} for parity and flavour symmetry, as well as the p.d.f. of
the topological charge density, are trivial in lattice QCD with
Ginsparg-Wilson fermions of non-vanishing mass. The essential
ingredient in what concerns the proof of parity realization in the
vacuum will be hermiticity. The rest of this article is organized as
follows. In section 2 we recall briefly the main features of
Ginsparg-Wilson fermions and analyze the one-flavour case. Section 3
generalizes the results of the previous section to an arbitrary number
of flavours, and shows how, contrary to the Wilson fermions
regularization, the two flavour model has not an Aoki phase in any
region of the parameter space. Section 4 contains a short discussion
on the expectations about the realization of these symmetries in the
chiral limit. In this limit the proof developed in the previous
sections does not apply, and therefore we will argue following the
standard wisdom. Hence, the content of that section is rather
speculative. Last section contains our conclusions.

\section{Ginsparg-Wilson fermions on the lattice}

Years ago Ginsparg and Wilson (G-W) \cite{Gins-Wil} suggested, in
order to avoid the Nielsen and Ninomiya non go theorem \cite{NoGo1,
  NoGo2} and to preserve chiral symmetry on the lattice, to require
the following condition for the inverse Dirac operator
\begin{equation}
\gamma_5 D^{-1} + D^{-1} \gamma_5 = 2aR \gamma_5,
\label{GW-I}
\end{equation}
where a is the lattice spacing and R is a local operator. Accordingly
D should satisfy, instead of the standard anticommutation relation of
the continuum formulation, the Ginsparg-Wilson relation
\begin{equation}
\gamma_5 D + D \gamma_5 = 2a D R \gamma_5 D.
\label{GW-II}
\end{equation}

Fifteen years after this proposal, Hasenfratz \cite{Hasen} and
Neuberger \cite{Neu1, Neu2} found that the fixed point action for QCD
and the overlap fermions satisfy respectively the Ginsparg-Wilson
relation, the last with R=1/2. Furthermore, Hasenfratz, Laliena and
Niedermayer \cite{VLaliena} realized that Ginsparg-Wilson fermions
have nice chiral properties, allowing us to establish an exact index
theorem on the lattice. Indeed if we define a local density of
topological charge as
\begin{equation}
q(x)= a R Tr\left(\gamma_5 D\left(x,x\right)\right)
\label{Topo-Q-I}
\end{equation}
one gets a topological charge 
\begin{equation}
Q = a R Tr(\gamma_5 D)
\label{Topo-Q-II}
\end{equation}
which is a topological invariant integer that approaches the continuum
topological charge in the continuum limit \cite{Kikukawa, Fujikawa,
  Suzuki, Adams1, Adams2}.

Finally by replacing the G-W Dirac operator $D+m$ by 
\begin{equation}
\Delta + m = \left(1-\frac{am}{2}\right)D + m
\label{GW-Op}
\end{equation}
in order to define an unsubtracted proper order parameter \cite{Chandra}
\begin{equation}
\bar\psi\left(1-\frac{aD}{2}\right)\psi,
\label{ChCon}
\end{equation}
it is easy to see \cite{Luscher} that the G-W fermionic action
possesses an exact symmetry which is anomalous for the flavour singlet
transformations, but exact for the flavour non-singlet case; a property
which allows us to introduce also a $\theta$ parameter in the G-W
action, as in the continuum.

In this paper we will work with G-W fermions that realize relation
\eqref{GW-II} with $R = 1/2$ (overlap fermions are a particular case
of that), and with the massive Dirac operator \eqref{GW-Op} associated
to the unsubtracted chiral order parameter \eqref{ChCon}.

We derive in appendix \ref{spectrum} some properties of the spectrum
of $\Delta + m$ and related operators that will be useful in the
following.

\section{The one flavour case}

Let us start by considering the one flavour case. We will make use of
the formalism developed in \cite{av, pdf}. First, we analyze the
standard bilinear \mbox{$O = i\bar\psi\gamma_5\psi$}. Our basic tool
is the function $P(q)$, which is the generating function for the
moments of $O$: $P(0) = 1$, $P'(0) = i <O>$, $\cdots$, $P^{n}(0) =
(i)^n <O^n>$, where the average is the lattice gauge theory average
with the full action. We can compute $P(q)$ as \cite{av}
\begin{equation}
P(q) = \langle 
\frac{\det(\Delta + m + \frac{q}{V} \gamma_5)}
{\det(\Delta + m)}
\rangle = 
\langle 
\frac{\det(H + \frac{q}{V})}
{\det(H)}
\rangle
\end{equation}
where V denotes the size of the matrix $\Delta$ (that is, up to a
constant, the number of points in the lattice), $H = \gamma_5
\left(\Delta + m \right)$, and the averages are now taken with the
effective gauge theory measure obtained after integrating out the
fermion fields
\begin{equation}
\left[ dA
\right ]
e^{-S_{YM}}{\det(\Delta + m)}
\end{equation}
We denote by $\mu_j$ the eigenvalues of $H$. Then we have for $P(q)$
\begin{equation}
P(q) = \langle 
\frac{\prod_j\left( \mu_j + \frac{q}{V} \right)}
{\prod_j(\mu_j)}
\rangle = 
 \langle 
\prod_j\left( 1 + \frac{q}{\mu_j V} \right)
\rangle
\end{equation}
In appendix \ref{spectrum} we show that $m \le |\mu_j|$, so everything
is well defined here as long as $m > 0$. Expanding the product we
rewrite $P(q)$ as
\begin{equation}
P(q) = \sum_{k = 0}^{V} q^k 
\langle 
\frac{1}{V^k} 
\sum_{\left(j_1, \cdots, j_k\right)} \frac{1}{\mu_{j_1}, \cdots, \mu_{j_k}}
\rangle
\label{pq}
\end{equation}
The sum in \eqref{pq} is taken over all different combinations of
k indices, so we can rearrange it as
\begin{equation}
\frac{1}{V^k}  
\sum_{\left(j_1, \cdots, j_k\right)} \frac{1}{\mu_{j_1}, \cdots,
  \mu_{j_k}} = 
\frac{1}{k!} \left(\frac{1}{V} \sum_j \frac{1}{\mu_j}\right)^k + 
\mathcal O(\frac{1}{V})
\label{pq2}
\end{equation}
But we show in appendix \ref{spectrum} that the $\mu$ come in pairs
$\pm \mu$, except maybe for the ones corresponding to chiral
modes. Therefore most of the terms in \eqref{pq2} cancel, and we are
left with the contribution coming from the chiral modes:
\begin{equation}
\frac{1}{V} \sum_j \frac{1}{\mu_j} = 
\frac{1}{V} \left( \frac{1}{m} \left( n^+ - n^-\right) +
\frac{a}{2}  \left( n'^+ - n'^-\right) \right)  = 
\left(\frac{am}{2} - 1 \right)
\frac{Q}{m V}
\end{equation}
Putting everything together, we have, for non-zero mass 
\begin{equation}
P(q) = \sum_{k = 0}^{V} \frac{q^k}{k!} 
\left(\frac{am}{2} - 1 \right)^k
\frac{1}{m^k}\langle\left(\frac{Q}{V}\right)^k\rangle + \mathcal O(\frac{1}{V})
\label{pq3}
\end{equation}

We see that in the thermodynamic limit the only dependence of $P(q)$
on the gauge field is through the topological charge. From \eqref{pq3}
the moments of the probability distribution of \mbox{$O =
  i\bar\psi\gamma_5\psi$} are related to the ones of the density of
topological charge $\frac{Q}{V}$ through
\begin{equation}
\langle \left(i\bar\psi\gamma_5\psi\right)^n\rangle = 
(-i)^n \left(\frac{am}{2} -1 \right)^n \frac{1}{m^n} 
\frac{\langle Q^n\rangle}{V^n}  + \mathcal O(\frac{1}{V})
\end{equation}

All the odd moments vanish by symmetry. The first non-trivial moment
is the second one
\begin{equation}
\langle\left(i\bar\psi\gamma_5\psi\right)^2\rangle = 
- \left(\frac{am}{2} - 1 \right)^2 \frac{1}{m^2} 
\langle\left(\frac{Q}{V}\right)^2\rangle + \mathcal O(\frac{1}{V})
\label{Parity}
\end{equation}
At this point we require that $i\bar\psi\gamma_5\psi$ be an hermitian
operator. This is a simple requirement, but with important
consequences: The expectation value of the square of an hermitian
operator must be positive, but from \eqref{Parity}, in the
thermodynamic limit it is manifestly negative. The only way to fulfill
both requirements at the same time is for the second moment to vanish,
\begin{equation}
\lim_{V\rightarrow\infty} \langle\left(\frac{Q}{V}\right)^2\rangle
= 0
\label{Q2}
\end{equation}
But then the probability distribution of the density of topological
charge, $Q/V$, must go to a delta at the origin
\begin{equation}
\lim_{V\rightarrow\infty} p\left(\frac{Q}{V}\right) =
\delta\left(\frac{Q}{V}\right)
\end{equation}
and all the higher moments of both $i\bar\psi\gamma_5\psi$ and
$\frac{Q}{V}$ vanish as well, therefore parity is not broken in
lattice QCD with one flavour of Ginsparg-Wilson fermions, at least for
the standard order parameter $i\bar\psi\gamma_5\psi$.

Let us consider now the case of the unsubstracted order parameter,
\begin{equation}
i \bar \psi \gamma_5 \left(1-\frac{aD}{2}\right) \psi
\end{equation}
We have 
\begin{equation}
P(q) = \langle 
\frac{\det(\Delta + m + 
\frac{q}{V} \gamma_5 \left(1 - \frac{a D}{2} \right))}
{\det(\Delta + m)}
\rangle 
\label{pq4}
\end{equation}
Let's prove that only the zero modes of $D$ contribute to
\eqref{pq4}. The matrix corresponding to the numerator is
block-diagonal, and the contribution to the ratio coming from a pair
of complex eigenvalues of $D$ is of the form (from \eqref{spectrum3} in
appendix \ref{spectrum})
\begin{equation}
1 - \alpha \frac{q^2}{V^2}
\end{equation}
with $|\alpha| \le m^{-2}$ \eqref{bound3}. Therefore the contribution
to $P(q)$ corresponding to complex eigenvalues comes from the factor
$\prod_j \left(1 - \alpha_j \frac{q^2}{V^2}\right)$, where the product
extends over all pairs of complex eigenvalues. Expanding this product,
we have for the coefficient corresponding to $q^{2k}$
\begin{equation}
\left|\frac{1}{V^{2k}}
\sum_{(j_1, \ldots, j_k)} \alpha_{j_1} \cdots \alpha_{j_k} \right|
\le \frac{1}{V^{2k}} \left(\sum_j |\alpha_j|\right)^k
\le \frac{V^k m^{-2k}}{V^{2k}} =  \frac{m^{-2k}}{V^k} 
\label{boundgeneral}
\end{equation}
Thus the contribution from the complex eigenvalues is of order $1 +
{\cal O}(\frac{1}{V})$, that is, just $1$ in the thermodynamic limit.

The chiral modes of $D$ with $\lambda = \frac{2}{a}$ contribute also
$1$ \eqref{spectrum4}. The zero modes of $D$, on the other hand,
give a non-trivial contribution:
\begin{equation}
P(q) = \left(1 + \frac{q}{mV}\right)^{n^+} 
\left(1 - \frac{q}{mV}\right)^{n^-} 
\end{equation}
If $n^+ < n^-$ ($Q > 0$),
\begin{equation}
P(q) = \left(1 - \frac{q^2}{m^2V^2}\right)^{n^+} 
\left(1 - \frac{q}{mV}\right)^Q 
\label{pq7}
\end{equation}
A similar expression is valid for $n^+ > n^-$. We can argue
essentially as in \eqref{boundgeneral} to see that the first factor in
\eqref{pq7} goes to $1$ in the thermodynamic limit. Therefore we
obtain the final result (valid for arbitrary values of $n^+$, $n^-$)
when $V \to \infty$
\begin{equation}
P(q) = \left(1 - {\mathrm sign}(Q) \frac{q}{mV}\right)^{|Q|}
\label{pq5} 
\end{equation}
All odd moments vanish as before because of (finite-volume) parity
symmetry. The even moments also vanish because they are trivially
related to the ones in \eqref{pq3}, as can be seen easily by expanding
\eqref{pq5}. In consequence we see that parity is not broken for the
unsubstracted order parameter either.

It is interesting to give a more physical argument that uses only the
vanishing of the second moment. In fact if parity were spontaneously
broken we would expect two degenerate vacua $\alpha$ and $\beta$,
since parity is a $Z_2$ symmetry. Let $z_\alpha$ be a complex number
which give us the mean value of the pseudoscalar $P$ in the $\alpha$
state
$$\left\langle P  \right\rangle_\alpha = z_\alpha,$$
then we have
$$\left\langle P  \right\rangle_\beta = - z_\alpha.$$
Since $P^2$ is parity invariant, it will take the same mean value in
the two states and making use of the cluster property in each one of
the two states we get
$$\left\langle P^2  \right\rangle = 
\frac{1}{2} \left\langle P^2  \right\rangle_\alpha + 
\frac{1}{2} \left\langle P^2  \right\rangle_\beta =
z_\alpha^2,$$
but since $\left\langle P^2 \right\rangle = 0$, $z_\alpha = 0$.

In conclusion we have shown rigorously, assuming hermiticity of $i
\bar\psi\gamma_5 \psi$ and using standard properties of
Ginsparg-Wilson fermions, that parity is not spontaneously broken in
the one-flavour model, at least for the standard order parameters.

\section{The $N_F$ flavours case}

We study now the case of $N_F$ flavours, in general with different,
non-zero masses. Most of the results from the previous section apply
here as well with small modifications. The fermionic action is
\begin{equation}
\sum_{\alpha=1}^{N_F} \bar\psi_\alpha \left(\Delta + m_\alpha\right) \psi_\alpha
\end{equation}
The complete spectrum of the Dirac operator consists of $N_F$ copies
of the single flavour spectrum, each of them calculated with the mass
of the corresponding flavour.

Let us consider the usual pseudoscalar order parameter for a single
flavour $\beta$, $i \bar\psi_\beta \gamma_5 \psi_\beta$. The
corresponding generating function $P(q)$ can be computed easily:
\begin{equation}
P(q) = \langle 
\frac{\det(\Delta + m_\beta + \frac{q}{V} \gamma_5)}
{\det(\Delta + m_\beta)}
\rangle_{N_F} 
\end{equation}
The average is taken over the effective gauge theory with $N_F$
flavours, but only the $\beta$ flavour appears within the
average\footnote{Our fermionic determinant (and hence, its
  eigenvalues) will always refer to a \emph{single flavour}.}. The
calculation is identical to the one for the single flavour case and
gives
\begin{equation}
P(q) = \langle \prod_j\left( 1 + \frac{1}{\mu_j^\beta} \frac{q}{V}
\right)\rangle
\label{pqnf}
\end{equation}
The superindex on the eigenvalues indicate flavour, that is,
$\mu^\beta$ belongs to the spectrum of $\gamma_5 \left(\Delta +
m_\beta \right)$. Equation \eqref{pqnf} is essentially the same as the
one for one flavour, and the moments of $P(q)$ still are, up to
constants, the same as the moments of the density of topological
charge $Q/V$.  The only difference is that now the average is taken in
the theory with $N_F$ flavours. This doesn't change any of the
conclusions: all odd moments vanish by symmetry, the second moment
vanish in the thermodynamic limit because of the hermiticity of $i
\bar\psi \gamma_5 \psi$, the density of topological charge goes to a
delta centered on the origin, and therefore all the higher moments
vanish as well (when $V \to \infty$). This is valid for any flavour
separately, and so will be valid as well for any linear combination
$\sum_\alpha A_\alpha i\bar\psi_\alpha\gamma_5\psi_\alpha$. The
extension to the unsubstracted order parameter is trivial.

Let us consider now the degenerate case, all flavours with equal
nonzero masses. The action now enjoys \emph{flavour symmetry}. As
stated in the introduction of this paper, the Vafa-Witten theorem
\cite{Vafa2} for vector-like symmetries does not apply to G-W fermions
because even if the integration measure is positive definite, the
Dirac operator does not anticommute with $\gamma_5$. However we can
study this symmetry with the p.d.f. method as we did for parity.

Consider first the case of two degenerate flavours and the standard
order parameters $\bar\psi \tau_3 \psi$ and $i \bar \psi \gamma_5
\tau_3 \psi$. Proceeding as before we find for the first order
parameter:
\begin{equation}
P(q) = \langle 
\frac{\det(\Delta + m + i \frac{q}{V})
\det(\Delta + m - i \frac{q}{V})}
{\det(\Delta + m)^2}
\rangle_{N_F} =
\langle \prod_j \left( 1 + \frac{q^2}{\lambda_j^2 V^2} \right) 
\rangle_{N_F}
\end{equation}
where $\lambda_j$ are the eigenvalues of $\Delta + m$.  By the same
argument we have use repeatedly before, $P(q) \to 1$ in the
thermodynamic limit\footnote{This conclusion rests only on the bound
  on the eigenvalues, and not on any other specific property of the
  Dirac operator.}.

Similarly for $i\bar\psi \gamma_5 \tau_3 \psi$ we obtain
\begin{equation}
P(q) = \langle \frac{\det(H + \frac{q}{V}) \det(H - \frac{q}{V})}
{\det(H^2)} \rangle_{N_F} = \langle \prod_j \left( 1 -
\frac{q^2}{\mu_j^2 V^2} \right) \rangle_{N_F}
\end{equation}
The same argument applies, and we have also that $P(q) \to 1$ in the
thermodynamic limit.

The calculations can be repeated easily for the unsubstracted
operators, and the result is the same. We can then conclude that here
is no Aoki phase for two flavours of Ginsparg-Wilson fermions with
non-zero mass. This conclusion should not be surprising at all, for
the spectrum of the Ginsparg-Wilson operator with a mass term is
depleted of small eigenvalues ($\approx \frac{1}{V}$).

The extension of this proof to $N_F$ flavours is trivial, because it
is easy to see that the preceding results apply to any pair of
flavours, and therefore general flavour symmetry is realized.

\section{The chiral limit}

The results of the previous sections can not be extended in a
straightforward way to QCD in the chiral limit. We lose the
non-trivial lower bound on the spectrum of $\Delta + m$, $0 < m <
|\lambda|$, and the Dirac operator has exact zero modes corresponding
to gauge fields with non-trivial topology.

As we can not get a definite conclusion on the realization of parity
in the chiral limit from first principles, we will follow in this case
the standard wisdom. Let us consider QCD with one massless flavour. In
this limit the action of the model has a new symmetry, the chiral U(1)
symmetry, which is anomalous because the integration measure is not
invariant under chiral U(1) global transformations. The Jacobian
associated to this change of variables introduce an extra term to the
pure gauge action proportional to the topological charge of the gauge
configuration, the $\theta$-vacuum term, which allows one to
understand the absence of a Goldstone boson in the model, but this
fact generates the well known strong CP problem. All these features of
QCD in the continuum formulation are well reproduced in lattice QCD
with Ginsparg-Wilson fermions, as discussed at the beginning of this
paper.

Let's consider first the unsubstracted scalar order parameter,
$\bar\psi\left(1 - \frac{aD}{2}\right)\psi$. The corresponding
generating function in the chiral limit and for a generic value of
$\theta$ can be written as
\begin{equation}
P(q) = \frac{\int\left[dA\right] e^{-S_{YM}}  e^{-i\theta Q}  
\det\left(D+i\frac{q}{V} \left(1 - \frac{aD}{2}\right)\right)}{Z}
\label{p0}
\end{equation}
with $Z = \int\left[dA\right] e^{-S_{YM}} e^{-i\theta Q}
\det\left(D\right)$ and $Q = n^- - n^+$. The contribution to the
determinant in the numerator of \eqref{p0} coming from pairs of
complex eigenvalues of $D$ is easily computed and gives a factor
\begin{equation}
f_0(q) = 
\prod_j \left(|\lambda_j|^2 - \frac{q^2}{V^2} 
\left( 1 - \frac{a^2|\lambda_j|^2}{4}\right)\right)
\end{equation}
where the product is taken over all different pairs of complex
eigenvalues. Each chiral mode corresponding to an eigenvalue
$\frac{2}{a}$ contributes a factor of $\frac{2}{a}$ to the
determinant, and each chiral mode corresponding to a zero eigenvalue
contributes a factor of $i \frac{q}{V}$. The normalization factor $Z$
is of course calculated from the same expressions by setting $q = 0$.
Therefore we can write $P(q)$ as
\begin{equation}
P(q) = Z^{-1} \int\left[dA\right] e^{-S_{YM}}  e^{-i\theta Q}  
\left(\frac{2}{a}\right)^{n'^+ + n'^-}
 \left(\frac{iq}{V}\right)^{n^+ + n^-}
f_0(q)
\label{ps}
\end{equation}

The computation for the pseudoscalar order parameter $i\bar\psi
\gamma_5\left(1 - \frac{aD}{2}\right)\psi$ is similar and the final
result is
\begin{equation}
P(q) = Z^{-1} \int\left[dA\right] e^{-S_{YM}}  e^{-i\theta Q}  
\left(\frac{2}{a}\right)^{n'^+ + n'^-}
\left(-1\right)^{n^+}
 \left(\frac{q}{V}\right)^{n^+ + n^-}
f_0(q)
\end{equation}
With the definition 
\begin{equation}
f_S(q) = \left(\frac{2}{a}\right)^{n'^+ + n'^-}
 \left(\frac{iq}{V}\right)^{n^+ + n^-}
f_0(q)
\end{equation}
and denoting by $P_S(q)$ and $P_P(q)$ the generating functions for the
scalar and pseudoscalar respectively, we can rewrite the above results
in the following way
\begin{eqnarray}
P_S(q) &=& Z_S^{-1} \int\left[dA\right] e^{-S_{YM}}  e^{-i\theta Q}  
f_S(q) \\
P_P(q) &=& Z_P^{-1} \int\left[dA\right] e^{-S_{YM}}  e^{-i\theta Q}  
f_S(q) (-i)^Q
\end{eqnarray}

From these expressions we see that only the $Q = \pm 1$ sectors
contribute to the chiral condensate \footnote{In fact only a subset of
  the $Q = \pm 1$ configurations give a non-vanishing contribution,
  those with only one zero mode, that is, that verify $n^- + n^+ = 1$
  as well as $n^- - n^+ = \pm 1$. Similar considerations apply to the
  other cases discussed in the text.} $\langle S \rangle$, and that
the dependence with the parameter $\theta$ is $\langle S \rangle
\propto \cos(\theta)$. For the second moment $\langle S^2\rangle$, the
only non-vanishing contributions come from the sectors with $Q = 0$
and $Q = \pm 2$, and we can write
\begin{equation}
\langle S^2\rangle = A_0 + A_2
\end{equation}
where $A_0$ is the contribution coming from the $Q = 0$ sector and
$A_2$ the contribution coming from the $Q = \pm 2$ sector. Looking at
the expression for the pseudoscalar condensate, all the odd moments
vanish because of parity symmetry. For the second moment we find that
\begin{equation}
\langle P^2\rangle = A_0 - A_2
\end{equation}
Looking at moments of higher order we would find an infinite set of
relations.

Since due to the chiral anomaly, strictly speaking, we do not have a
new symmetry in the chiral limit of one flavour QCD, the standard
wisdom is that the vacuum expectation value of the chiral order
parameter will not vanish as $m\rightarrow 0$. Moreover, the absence
of massless particles in the one flavour model suggests that the
perturbation series in powers of $m$ does not give rise to infrared
divergences \cite{LeutSmilg}, the free energy density is an ordinary
Taylor series in $m$ \cite{Niedermayer,LeutSmilg}; and in what
concerns the chiral condensate, the chiral and thermodynamical limits
commute.

On the other hand, the free energy density of the model at $m\ne 0$
and $\theta=0$ can be computed in the thermodynamical limit from the
topologically trivial sector $Q=0$ \cite{Niedermayer,LeutSmilg}. But
since chiral symmetry in the $Q=0$ sector is not anomalous, it should
be spontaneously broken in the topologically trivial sector if the
chiral condensate takes a non vanishing value when approaching the
chiral limit. In such a case, the value of the chiral condensate in
the full theory and in the chiral limit will be related to the
spectral density at the origin of eigenvalues of the Dirac operator of
the topologically trivial sector by the well known Banks and Casher
formula \cite{BanksCasher}
$$\left\langle S\right\rangle = -\pi\rho(0) = \Sigma_0.$$
This equation provide us with a non trivial relation between the value
of the scalar condensate in the chiral limit in the full theory, which
gets all its contribution from the $Q=+\- 1$ sectors, and the
spectral density of eigenvalues at the origin of the Dirac operator
in the topologically trivial sector $Q=0$.

The scalar condensate is invariant under parity, and therefore in the
full theory in the chiral limit, irrespective of the realization of
parity in the vacuum we expect its probability distribution function
to be a delta function $\delta(c-\Sigma_0)$. Therefore, we expect for
the moments
\begin{equation}
\langle S^n \rangle = \Sigma_0^n
\end{equation}
For the second moment, this tells us that
\begin{equation}
A_0 + A_2 = \Sigma_0^2
\end{equation}
But as previously stated, the standard wisdom tell us that the
topologically trivial sector breaks spontaneously the chiral $U(1)$
symmetry, and since $A$ is exactly the second moment of
$P_{\mathcal{S}}(c)$ computed in the $Q=0$ sector, we should have
\begin{equation}
A_0 = \frac{1}{2} \Sigma_0^2
\end{equation}
and therefore also 
\begin{equation}
A_2 = \frac{1}{2} \Sigma_0^2
\end{equation}
But this implies that the second moment of the pseudoscalar condensate
vanishes,
\begin{equation}
\langle P^2 \rangle = 0
\end{equation}
By a similar argument we can see that for higher (even) moments,
$\langle S^{2n} \rangle = \Sigma_0^{2n}$ implies the vanishing of the
corresponding pseudoscalar moment, $\langle P^{2n} \rangle = 0$

Symmetry under parity is the only obvious reason for the vanishing of
the pseudoscalar moments, and therefore the previous result strongly
suggests that parity is also realized in QCD with one massless flavour.

\section{Conclusion}

Although the common lore on QCD symmetries states that parity and
vector-like global symmetries remain unbroken, no sound theoretical
proof of this hypothesis has ever been presented. The arguments given
by Vafa and Witten against spontaneous breaking of these symmetries in
\cite{Vafa1,Vafa2} were questioned by several groups
\cite{ag,Ji,Crom,av}, and now there is agreement in the scientific
community on the lack of a proof for parity realization in
QCD. Concerning vector-like symmetries as flavour or baryon number
conservation, it must be remarked that the staggered fermion
discretization is the only known lattice regularization that fulfills
the initial conditions of the Vafa-Witten theorem. Indeed the theorem
is not applicable neither to the Ginsparg-Wilson regularization nor to
Wilson's one, widely used for lattice QCD simulations. In the first
case the theorem does not apply because even if the integration
measure is positive definite, the Dirac operator does not anticommute
with $\gamma_5$. In the second case neither of the two assumptions,
positivity of the integration measure and anticommutation of the Dirac
operator with $\gamma_5$ are fulfilled. Indeed, there exists for
Wilson fermions a region of parameter space where parity and flavour
symmetries are spontaneously broken, the well known Aoki phase.

On the other hand the p.d.f. formalism can be used to cast some light
on the old aim of understanding the realization of symmetries of QCD
from first principles. In fact, we have shown in this paper that some
interesting conclusions appear when we apply the p.d.f. formalism to
the Ginsparg-Wilson regularization. There, we see how the more
standard order parameters for parity and flavour symmetries take a
vanishing vacuum expectation value for a non-zero fermion mass. This
is a major result that overcomes the difficulties found by
\cite{Vafa1,Vafa2}.

\section{Acknowledgments}

This work has been partially supported by an INFN-MEC collaboration,
CICYT (grant FPA2009-09638) and DGIID-DGA
(grant2007-E24/2). E. Follana is supported by Ministerio de Ciencia e
Innovaci\'on through the Ram\'on y Cajal program, and A. Vaquero is
supported by the Ministerio de Educaci\'on through the FPU program.

\appendix

\section{Spectrum of D and Related Operators}
\label{spectrum}

We start with the Ginsparg-Wilson relation with $R = 1/2$,
\begin{equation}
\gamma_5 D + D \gamma_5 = a D \gamma_5 D.
\label{GW-II-A}
\end{equation}
We can also choose $D$ such that 
\begin{equation}
\gamma_5D\gamma_5 = D^\dagger
\label{ConjHerm}
\end{equation}
From \eqref{GW-II-A} and \eqref{ConjHerm} we obtain
\begin{equation}
D + D^\dagger = a D D^\dagger = a D^\dagger D
\label{normal}
\end{equation}
Therefore $D$ is a normal operator, and as such has a basis of
orthonormal eigenvectors. Also eigenvectors corresponding to different
eigenvalues are necessarily orthogonal. From \eqref{normal} it is
immediate to check that the operator $V = 1-aD$ is unitary, $V^\dagger
V = I$. Therefore the spectrum of $V$ lies in the unit circle with
center in the origin, and the spectrum of $D$ must then lie in the
shifted and rescaled circle of radius $\frac{1}{a}$ centered in the
real axis at $(\frac{1}{a},0)$. The possible eigenvalues of $D$ are
therefore of the form
\begin{equation}
\lambda = \frac{1}{a} \left(1 - e^{i \alpha} \right), \alpha \in
\mathbb{R}
\label{eigen}
\end{equation}
We also have the identity 
\begin{equation}
\lambda + \lambda^* = a \lambda \lambda^*
\label{circle}
\end{equation}

Let $\mathbf{v}$ be an eigenvector of $D$ with eigenvalue $\lambda$,
$D\mathbf{v} = \lambda\mathbf{v}$. Taking into account \eqref{GW-II-A}
\begin{equation}
D\gamma_5\mathbf{v} = -\gamma_5D\mathbf{v} + aD\gamma_5D\mathbf{v} =
-\lambda\left(\gamma_5\mathbf{v} + aD\gamma_5\mathbf{v}\right)
\label{Eigen-I-A}
\end{equation}
Therefore using \eqref{eigen}
\begin{equation}
D \left(\gamma_5\mathbf{v}\right) =
-\frac{\lambda}{1-a\lambda} \left(\gamma_5\mathbf{v}\right) = 
\lambda^{*}\left(\gamma_5\mathbf{v}\right).
\end{equation}
Thus, if $\mathbf{v}$ is an eigenvector of $D$ with eigenvalue
$\lambda$, then $\gamma_5\mathbf{v}$ is another eigenvector with
eigenvalue $\lambda^{*}$, and if $\lambda$ is not real then those two
eigenvectors correspond to different eigenvalues and must be
orthogonal. On the other hand if we restrict to the subspace
corresponding to real eigenvalues, $\lambda = 0$ or $\lambda =
\frac{2}{a}$, $\gamma_5$ and $D$ commute, and therefore we can find a
common basis of eigenvectors; in other words, we can find an
orthonormal basis for which the eigenvectors of $D$ corresponding to
real eigenvalues are chiral. If we denote by $n^+$ ($n^-$) the number
of eigenvectors of positive (negative) chirality in the subspace
corresponding to $\lambda = 0$, and similarly $n'^+$ ($n'^-$) for the
subspace corresponding to $\lambda = \frac{2}{a}$, then $Tr(\gamma_5)
= 0$ and $Q = \frac{a}{2} Tr(\gamma_5 D)$ imply
\begin{eqnarray}
n^+ - n^- &=& n'^- - n'^+ \\
Q &=& n^- - n^+
\end{eqnarray}
We denote by $V$ the size of the matrix $D$. Then the density of
topological charge, defined as $\frac{Q}{V}$, is bounded in absolute
value by 1, $\left|\frac{Q}{V}\right| \le 1$.

The operator in the fermion action is
\begin{equation}
\Delta + m = \left(1-\frac{am}{2}\right)D + m
\label{GW-Op-A}
\end{equation}
Its spectrum is trivially related to the spectrum of $D$; if $\lambda$
are as before the eigenvalues of $D$, then the eigenvalues of
\eqref{GW-Op-A} are $\left( 1 - \frac{am}{2}\right) \lambda + m$. They
still lie in a circle with the center in the real axis, and the
possible real eigenvalues are now $m$ and $\frac{2}{a}$. We will
always require that $ 0 < m < \frac{2}{a}$, then the operator
\eqref{GW-Op-A} preserves the position of the higher real eigenvalue
\cite{Niedermayer}.

We will also need the spectrum of $ H =
\gamma_5\left(\Delta+m\right)$. It is easy to see that $H$ is an
hermitian operator, $H^\dagger = H$, and therefore has real spectrum
$\mu_j$. We can calculate this spectrum by noting that the matrix
$\gamma_5\left(\Delta+m\right)$ is block diagonal in the basis of
eigenvectors of $D$. If we denote by $\mathbf{v}_{\lambda}$ such and
eigenvector with non-real eigenvalue $\lambda$, we have
\begin{eqnarray}
H \mathbf{v}_\lambda = 
\gamma_5 \left(1 - \frac{am}{2}\right) \lambda \mathbf{v}_\lambda
+ m  \mathbf{v}_{\lambda^{*}} = 
\left(m + \lambda \left( 1 - \frac{am}{2}\right)\right)
\mathbf{v}_{\lambda^{*}}
\\
H \mathbf{v}_{\lambda^{*}} = 
\gamma_5 \left(1 - \frac{am}{2}\right) \lambda^{*} \mathbf{v}_{\lambda^{*}}
+ m  \mathbf{v}_{\lambda} = 
\left(m + \lambda^{*} \left( 1 - \frac{am}{2}\right)\right)
\mathbf{v}_{\lambda}
\end{eqnarray}
We therefore have a $2\times 2$ block
\begin{equation}
\left(\begin{array}{cc}
0 & m + \lambda \left(1 - \frac{am}{2} \right)\\
 m + \lambda^* \left(1 - \frac{am}{2}\right) & 0 
\end{array}
\right)
\end{equation}
The diagonalization of this block yields a pair of real eigenvalues $\pm
\mu$ with
\begin{equation}
\mu^2 = m^2 + \lambda \lambda^* \left(1 - \frac{am}{2} \right)^2
+ m \left(\lambda + \lambda^*\right) \left(1 - \frac{am}{2}\right)
\end{equation}

For $\lambda$ real, let $\mathbf{v}_\lambda$ be an eigenvector of $D$
of chirality $\chi$, that is, $\gamma_5 \mathbf{v}_\lambda =
\chi \mathbf{v}_\lambda$. Then
\begin{equation}
H \mathbf{v}_\lambda = 
\gamma_5 \left(1 - \frac{am}{2}\right) \lambda \mathbf{v}_\lambda
+ m  \chi \mathbf{v}_\lambda = 
\left(m + \lambda \left( 1 - \frac{am}{2}\right) \right)
\chi \mathbf{v}_\lambda
\end{equation}
That is,
\begin{equation}
\mu = 
\left(m + \lambda \left( 1 - \frac{am}{2}\right) \right)
\chi 
\end{equation}
More explicitly, we have $\mu = m$ and $\mu = - m$ with degeneracy
$n^+$ and $n^-$ respectively, and similarly $\mu = \frac{a}{2}$, $\mu
= - \frac{a}{2}$ with degeneracy $n'^+$, $n'^-$.

Note that $\det(\gamma_5 (\Delta + m) = \det(\Delta + m)$, and
therefore 
\begin{equation}
\prod_j \lambda_j = \prod_j \mu_j
\end{equation}
From the above calculation we also obtain immediately a bound for
$\mu$ at finite mass (this was remarked in \cite{Niedermayer}):
\begin{equation}
\mu^2 \ge m^2
\end{equation}

Let's consider now the operator 
\begin{equation}
\Delta + m + 
\frac{q}{V} \gamma_5 \left( 1 - \frac{aD}{2}\right)
\end{equation}
We are interested in its determinant. Proceeding as before, we see
that it is also block-diagonal in the basis of eigenvectors of $D$,
and the contribution to the determinant coming from the block
corresponding to a complex pair $\mathbf{v}_\lambda$,
$\mathbf{v}_{\bar\lambda}$ is given by 
\begin{multline}
\det{\left(\begin{array}{cc}
\left(1 - \frac{am}{2}\right) \lambda + m & 
\frac{q}{V} \left(1 - \frac{a \lambda}{2} \right) 
\\
\frac{q}{V} \left(1 - \frac{a \lambda^*}{2} \right) &
\left(1 - \frac{am}{2}\right) \lambda^* + m
\end{array}
\right)} = \\
m^2 + 
\left[ 1 - \left(\frac{am}{2}\right)^2\right] \lambda \lambda^* -
\frac{q^2}{V^2} \left( 1 - \frac{a^2 \lambda \lambda^*}{4} \right)
\label{spectrum3}
\end{multline}
where we have used identity \eqref{circle}.  We also have the bound
\begin{equation}
\left|
\frac{1 - \frac{a^2 \lambda \lambda^*}{4}}
{m^2 + \left( 1 - \frac{(am)^2}{4}\right) \lambda \lambda^*}
\right|  \le \frac{1}{m^2}
\label{bound3}
\end{equation}

A chiral mode $\lambda$ with chirality $\chi$  gives a contribution
\begin{equation}
\left(1 - \frac{am}{2} \right) \lambda + m +
\frac{q}{V} \left( 1 - \frac{a \lambda}{2}\right) \chi 
\label{spectrum4}
\end{equation}
For $\lambda = \frac{2}{a}$ the contribution is just $\frac{2}{a}$,
whereas for a zero mode $\lambda = 0$ with chirality $\chi$, the
contribution is $m + \chi \frac{q}{V}$.

\end{document}